\documentclass{PoS}

\title{
    \begin{picture}(0,0)(0,0)%
    \put(350,55){\makebox(0,0)[l]{\textnormal{\normalsize RIKEN-QHP-254}}}%
    \end{picture}
Baryon interactions in lattice QCD: \\ the direct method vs. the HAL QCD potential method}

\ShortTitle{Baryon interactions in lattice QCD}

\author{\speaker{Takumi Iritani}\\
  Department of Physics and Astronomy,
  Stony Brook University, NY 11794-3800, USA, and \\
  Theoretical Research Division, Nishina Center, RIKEN, Wako 351-0198, Japan \\
  E-mail: \email{takumi.iritani@stonybrook.edu}}

\author{for HAL QCD Collaboration\\
  \includegraphics[width=0.35\textwidth]{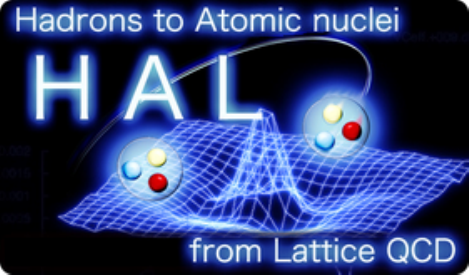}
}

\abstract{
  We make a detailed comparison between the direct method and the HAL QCD potential method
  for the baryon-baryon interactions, 
  taking the $\Xi\Xi$ system at $m_\pi= 0.51$ GeV in 2+1 flavor QCD and
  using  both smeared and wall quark sources.
  The energy shift $\Delta E_\mathrm{eff}(t)$ in the direct method shows 
  the strong  dependence on the choice of quark source operators,
  which means  that the results with either (or both) source are false.
  The time-dependent HAL QCD method,  on the other hand, 
  gives the quark source independent $\Xi\Xi$ potential, thanks to the
  derivative expansion of the potential, which absorbs the source dependence to the next leading order correction.
  The HAL QCD potential predicts the absence of the bound state in the $\Xi\Xi$($^1$S$_0$) channel 
  at $m_\pi= 0.51$ GeV, which is also confirmed by the volume dependence of finite volume energy from the potential.
  We also demonstrate that
  the origin of the fake plateau in the effective energy shift $\Delta E_\mathrm{eff}(t)$
  at $t \sim 1$ fm can be clarified by
  a few low-lying eigenfunctions and eigenvalues on the finite volume derived from the HAL QCD potential,
  which implies that the ground state saturation of $\Xi\Xi$($^1$S$_0$) 
  requires $t \sim 10$ fm in the direct method for the smeared source on $(4.3 \ \mathrm{fm})^3$ lattice,
  while the HAL QCD method does not suffer from such a problem.
}

\FullConference{34th annual International Symposium on Lattice Field Theory\\
		24-30 July 2016\\
		University of Southampton, UK}

\begin{document}

\section{Introduction}
  \vspace{-2.00ex}
Although L\"uscher's finite volume method \cite{Luscher:1991}
and HAL QCD method \cite{Ishii:2006ec}
are theoretically equivalent and employed to study hadron-hadron interactions in lattice QCD
\cite{FVM-review,Aoki:2012tk,Kurth:2013tua,Yamazaki:2012hi,Yamazaki:2015asa,Beane:2011iw,
HALQCD:2012aa,Charron:2013paa,Yamada:2015cra},
two methods give inconsistent results for two-baryon systems (see a review \cite{FVM-review}).

Recently, we pointed out that the direct measurement of 
the two-baryon energy shift in L\"uscher's method suffers from systematic uncertainties
due to contamination of excited states \cite{Iritani:2016jie,Iritani:2015dhu} that
plateaux in the effective energy shift $\Delta E_\mathrm{eff}(t)$ disagree between
smeared and wall sources.
In this talk, we clarify the origin of the fake plateaux in the direct method 
using the HAL QCD potential, which is insensitive to source operators. 

  \vspace{-2.00ex}
\section{Formalism}
  \vspace{-2.00ex}
\subsection{L\"uscher's finite volume method}

The energy shift of the two-body system in the finite volume $L$, 
$\Delta E_\mathrm{BB}(L) = E_\mathrm{BB}(L) - 2m_B$, with the ground state energy
of the two-baryon system $E_\mathrm{BB}(L)$ and a single baryon mass $m_B$, is related to
the phase shift $\delta(k)$ through the finite volume formula \cite{Luscher:1991} as
\begin{equation}
  k\cot\delta(k) = \frac{1}{\pi L} \sum_{\vec{n}\in\mathbf{Z}^3}
  \frac{1}{|\vec{n}|^2-|kL/(2\pi)|^2}, 
  \label{eq:Luscher}
\end{equation}
where $k$ is defined by $\Delta E_\mathrm{BB}(L) = 2 \sqrt{(m_B)^2 + k^2} - 2m_B$. 
The bound state is determined from the pole condition, $k\cot\delta(k) = -
\sqrt{-k^2}$ at $L \rightarrow \infty$
\footnote{A systematic diagnosis of the phase shift of the previous studies 
  \cite{Yamazaki:2012hi,Yamazaki:2015asa,Beane:2011iw}
is discussed in Ref.~\cite{Aoki:2016}}.

In lattice QCD simulations, $\Delta E_\mathrm{BB}(L)$ is estimated by the plateau of the
effective energy shift 
\begin{equation}
  \Delta E_\mathrm{BB}^\mathrm{eff}(t) \equiv
  E_\mathrm{BB}^\mathrm{eff}(t) - 2m_\mathrm{B}^\mathrm{eff}(t)
  = - \frac{1}{a} \log \left( \frac{R_\mathrm{BB}(t+a)}{R_\mathrm{BB}(t)} \right), 
  \label{}
\end{equation}
where $R_\mathrm{BB}(t) \equiv C_\mathrm{BB}(t)/\{C_\mathrm{B}(t)\}^2$
with the two-baryon propagator $C_\mathrm{BB}(t) \equiv \left\langle B(t)^2
\bar{B}(0)^2 \right\rangle$,
the baryon propagator $C_B(t) \equiv \left\langle B(t)\bar{B}(0) \right\rangle$
and the lattice spacing $a$. 

  \vspace{-1.00ex}
\subsection{Difficulties in multi-baryon systems}
  \vspace{-1.00ex}

Besides its significant computational cost, 
the multi-baryon systems in lattice QCD
has the signal to noise ratio problem,
which becomes exponentially worse for $A$ baryons as 
$S(t)/N(t) \sim \exp\left[ - A(m_B - (3/2)m_M)t \right]$,
where $m_B$ and $m_M$ are the baryon and meson masses.
In addition to this, the direct method suffers from the contamination of elastic scattering states,
whose energy gap decrease as $\mathcal{O}(1/L^2)$ as the volume increases.
For example, 
a gap between the ground state and the first $\Xi\Xi$ scattering state is about
$50$ MeV at $L=4.3\ \mathrm{fm}$ in this study, which
requires $ (50\ \mathrm{MeV})^{-1} \ll t \sim \mathcal{O}(10)$ fm for the ground state saturation.

\begin{figure}[h]
  \centering
  \includegraphics[width=0.47\textwidth,clip]{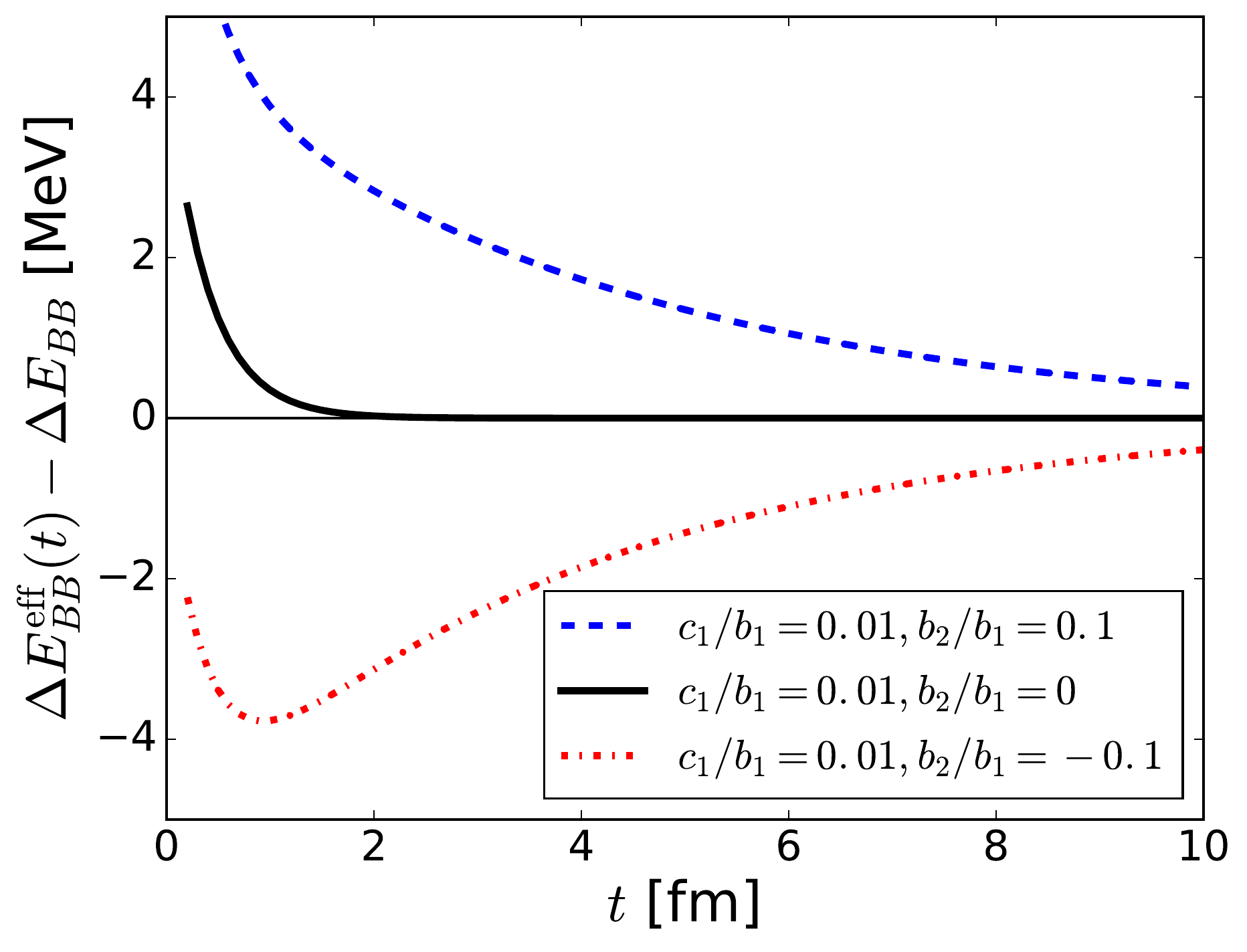}
  \includegraphics[width=0.47\textwidth,clip]{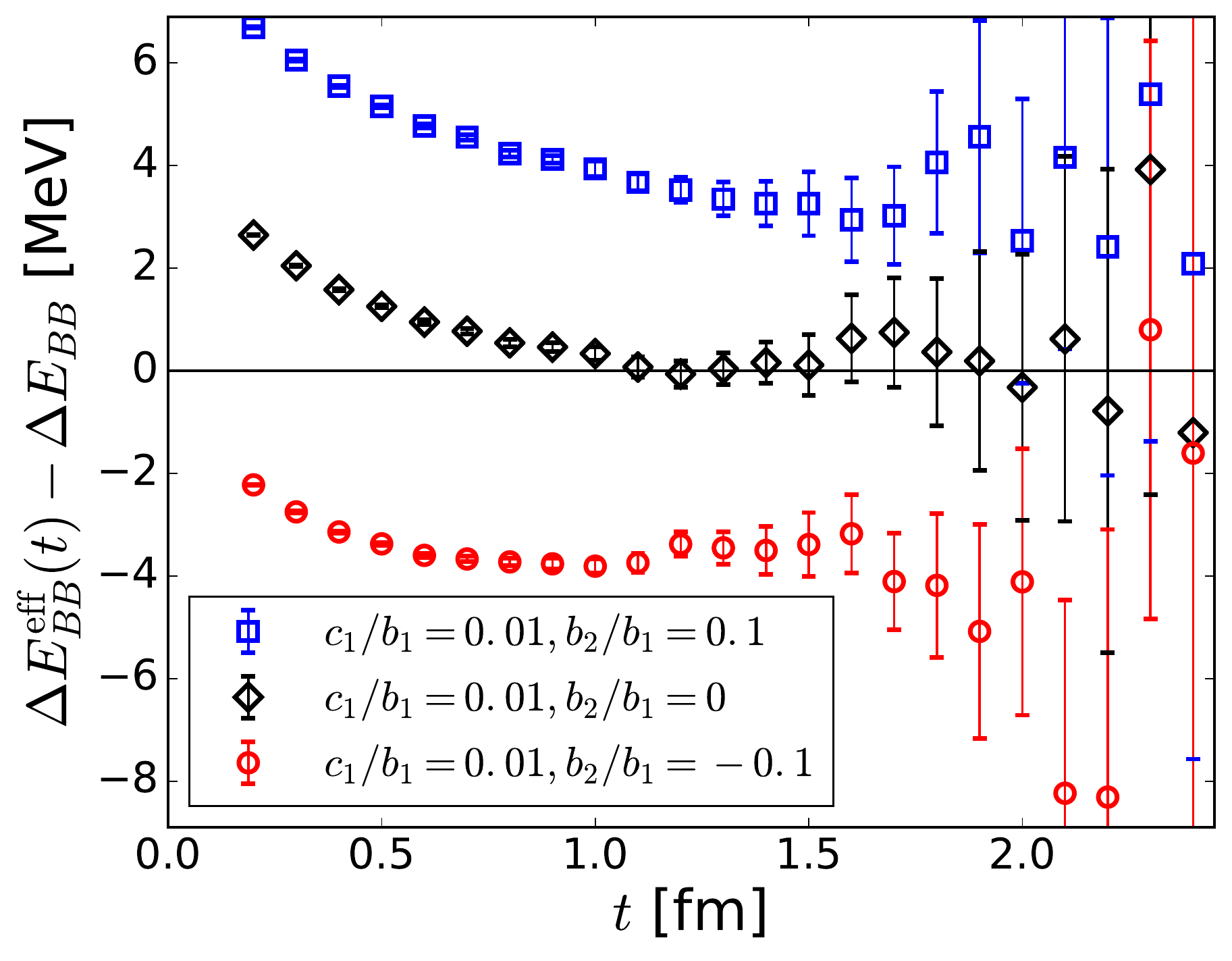}
  \caption{(a) The effective energy shift of the mock-up data. 
    (b) A mock-up data with fluctuations.
  \label{fig:mock}}
\vspace{-2ex}
\end{figure}

As an instructive example \cite{Iritani:2016jie} , let us consider the mock-up data as
\begin{equation}
  R(t) = b_1 e^{-\Delta E_\mathrm{BB}t} + b_2 e^{-(\delta E_\mathrm{el}+\Delta E_\mathrm{BB})t}
  +c_1 e^{-(\delta E_\mathrm{inel}+\Delta E_\mathrm{BB})t},
  \label{}
\end{equation}
where $\Delta E_\mathrm{BB} = E_\mathrm{BB} - 2m_\mathrm{B}$,
while 
$\delta E_\mathrm{el} = E_\mathrm{BB}^\ast - E_\mathrm{BB}$
and $\delta E_\mathrm{inel} = E_\mathrm{inel} - E_\mathrm{BB}$ for the excited states.
Fig.~\ref{fig:mock}(a) shows 
the lines of $\Delta E_\mathrm{BB}^\mathrm{eff}(t) - \Delta E_\mathrm{BB}$
at $\delta E_\mathrm{el} = 50$ MeV and $\delta E_\mathrm{inel} = 500$ MeV, 
which are typical values for the elastic and inelastic excitations, with
$c_1/b_1 = 0.01$ and $b_2/b_1 = \pm 0.1$, $0$.
Without the elastic state ($b_2/b_1 = 0$),
$\Delta E_\mathrm{BB}^\mathrm{eff}(t)$ converges to $\Delta E_\mathrm{BB}$
around $t \sim 1$ fm within 1 MeV of accuracy, while
the ground state saturation requires 
$t \sim 10$ fm even for the 10\% contamination.

Figure~\ref{fig:mock}(b) is the discrete data with fluctuations added.
There appear plateau-like structures around $t \sim 1$ fm,
which however are fake plateaux as seen in Fig.~\ref{fig:mock}(a).
This demonstrates a difficulty in obtaining
the ground state energy from 
a plateau-like structure in $\Delta E_\mathrm{eff}(t)$ at $t\simeq 1$ fm.

  \vspace{-1.00ex}
\subsection{HAL QCD method}
  \vspace{-1.00ex}
Contrary to the direct method,
the time-dependent HAL QCD method \cite{HALQCD:2012aa} utilizes all scattering state below the inelastic threshold 
to extract the non-local potential $U(\vec{r}, \vec{r}^\prime)$ as
\begin{equation}
  \left[ \frac{1}{4m_B}\frac{\partial^2}{\partial t^2}
  - \frac{\partial}{\partial t} - H_0\right]R(\vec{r},t)
  = \int d\vec{r}^\prime U(\vec{r},\vec{r}^\prime)R(\vec{r}^\prime,t)
  \label{}
\end{equation}
for $t \gg (\Delta W_\mathrm{th})^{-1}$, where  the Nambu-Bethe-Salpeter (NBS) correlation function $R(\vec{r},t)$ is defined as
\begin{equation}
  R(\vec{r},t) \equiv 
  \left\langle 0 | T\{B(\vec{x}+\vec{r},t)B(\vec{x},t)\bar{\mathcal{J}}(0) | 0 \right\rangle
    /\{C_B(t)\}^2
    = \sum_n A_n \phi^{W_n}(\vec{r}) e^{-\Delta W_n t}
    + \mathcal{O}(e^{-\Delta W_\mathrm{th}t})
  \label{}
\end{equation}
with a source operator $\mathcal{J}$,
$\Delta W_n = W_n - 2m_B$ with $n$-th energy eigenvalue $W_n$,
and the inelastic threshold $\Delta W_\mathrm{th} = W_\mathrm{th} - 2m_B$.
Using the velocity expansion, $U(\vec{r},\vec{r}^\prime) \simeq
\{V(\vec{r}) + \mathcal{O}(\nabla^2)\}
\delta(\vec{r}-\vec{r}^\prime)$,
the leading order potential is given by
\begin{equation}
  V(\vec{r}) = \frac{1}{4m_B} \frac{(\partial/\partial t)^2 R(\vec{r},t)}{R(\vec{r},t)}
  - \frac{(\partial/\partial t)R(\vec{r},t)}{R(\vec{r},t)}
  - \frac{H_0 R(\vec{r},t)}{R(\vec{r},t)}.
  \label{eq:HAL_LO}
\end{equation}

  \vspace{-1.00ex}
\section{Lattice QCD measurements for $\Xi\Xi$ interactions}
  \vspace{-1.00ex}

We use 2+1 flavor QCD ensembles in  Ref.~\cite{Yamazaki:2012hi}, generated with the Iwasaki gauge action
and $\mathcal{O}(a)$-improved Wilson quark action at 
$a = 0.8995(40)$ fm, where $m_\pi = 0.51$ GeV, $m_N = 1.32$ GeV and $m_\Xi = 1.46$ GeV.
For a comparison, we employ the wall source $q^\mathrm{wall}(t) = \sum_{\vec{y}} q(\vec{y},t)$,
which is mainly used in the HAL QCD method,
as well as the smeared source $q^\mathrm{smear}(\vec{x},t) =
\sum_{\vec{y}} f(|\vec{x}-\vec{y}|)q(\vec{y},t)$
with $f(r) \equiv Ae^{-Br}, 1, 0$ for $0 < r < (L-1)/2$, $r = 0$, $(L-1)/2 \leq r$, which is
generally adopted for the direct method.
Simulation parameters including $(A,B)$ identical to those in Ref.~\cite{Yamazaki:2012hi}
are summarized in Table~\ref{tab:lat_param}.
In this report, we mainly consider $\Xi\Xi$($^1$S$_0$)  channel using the relativistic interpolating operators, since $\Xi\Xi$($^1$S$_0$) channel  has smaller statistical errors  but belongs to  
the same SU(3) flavor representation of the NN($^1$S$_0$).

  \vspace{-1.00ex}
  \subsection{Quark source dependence of the effective energy shift $\Delta E_{\Xi\Xi}^\mathrm{eff}(t)$}
  \vspace{-1.00ex}

\begin{figure}[h]
  \centering
  \includegraphics[width=0.47\textwidth,clip]{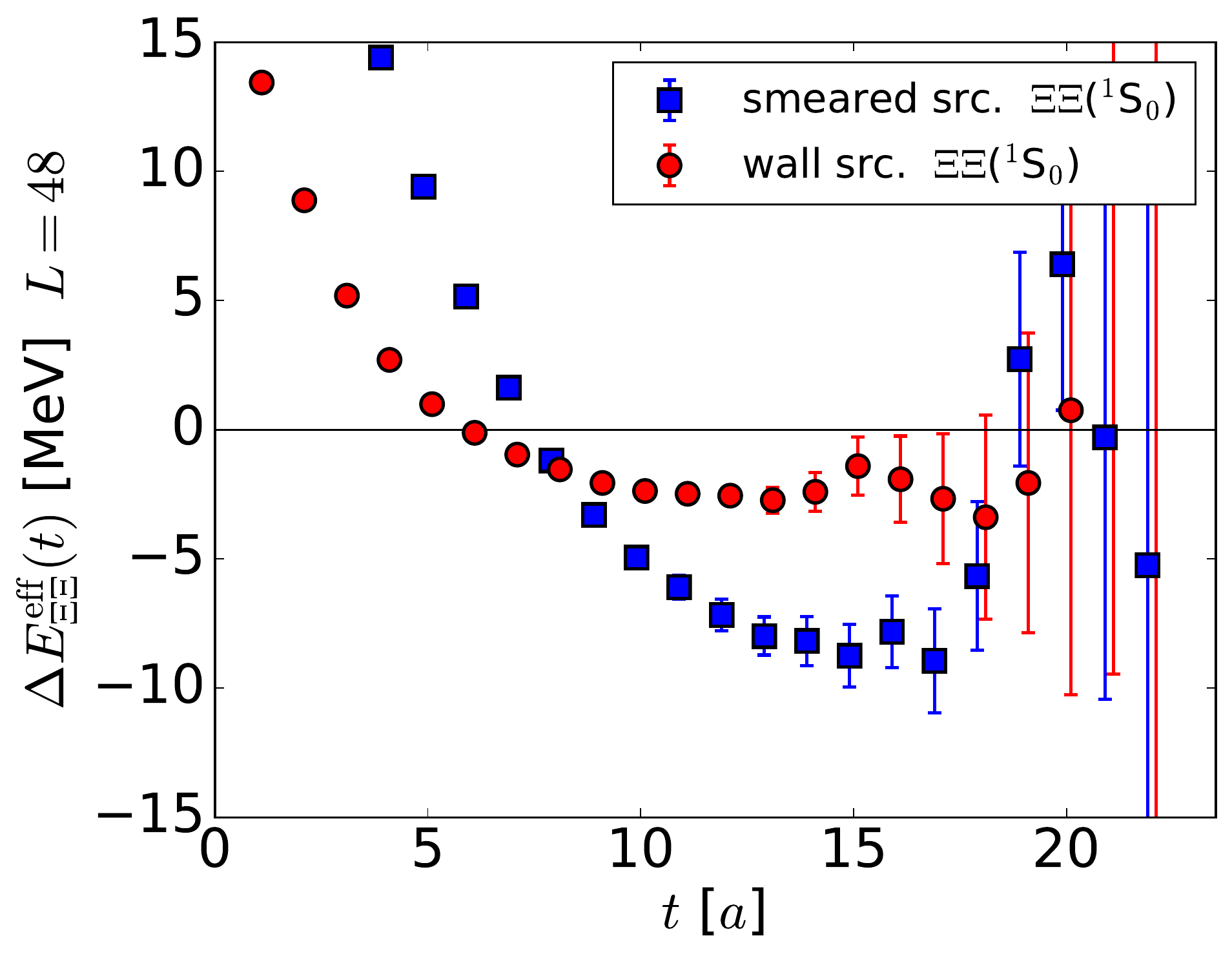}
  \includegraphics[width=0.47\textwidth,clip]{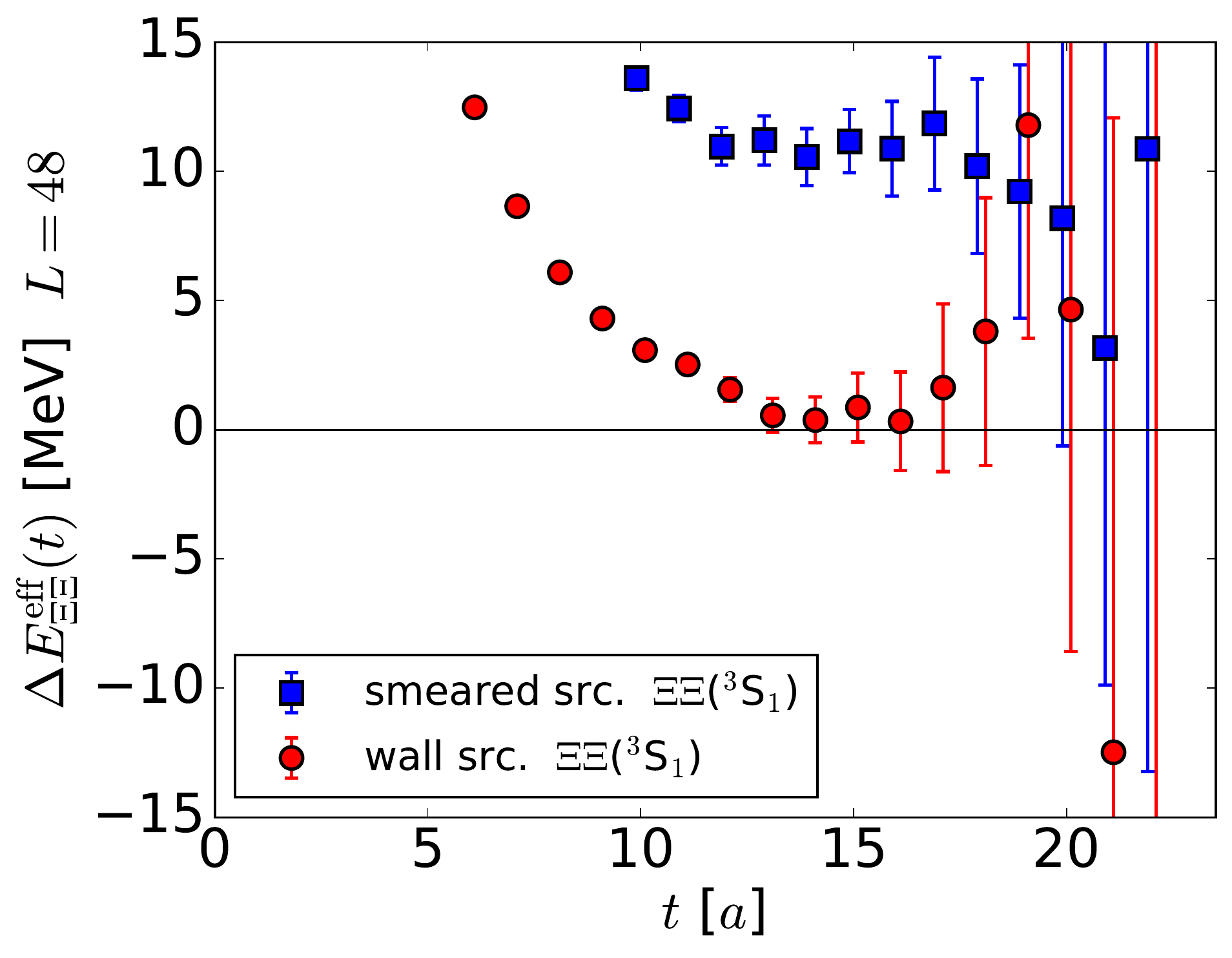}
  \caption{Examples of the effective energy shift plots at $L^3 = 48^3$. \label{fig:EeffSample}}
\vspace{-2ex}
\end{figure}

Quark source dependence is an easy check against fake plateaux.
We compare the effective energy shift between the wall and smeared sources
in Fig.~\ref{fig:EeffSample} for $\Xi\Xi$($^1$S$_0$) (Left) and $\Xi\Xi$($^3$S$_1$) (Right) 
on $48^3$ lattice.
While plateau-like structures appear around $t = 15a$ for both sources,
they disagree with each other,
implying that either plateau (or both) is fake.
Repeating this analysis on other volumes and taking $L\rightarrow\infty$ limit,
we have found that 
the lowest energy state from the wall source is the scattering state in both
$\Xi\Xi$($^1$S$_0$) and $\Xi\Xi$($^3$S$_1$) channels, while
that from the smeared source turns out to be the bound state in the $\Xi\Xi$($^1$S$_0$) channel but
an unphysical state in the $\Xi\Xi$($^3$S$_1$),  which has positive energy
shift $\Delta E_\mathrm{\Xi\Xi}(^3$S$_1) > 0$ in the infinite volume limit.
These results bring serious doubt on the validity of the energy shift 
in the previous works \cite{Yamazaki:2012hi,Yamazaki:2015asa,Beane:2011iw}
\footnote{The possibility of the fake plateau can be checked by the finite volume formula ~\cite{Aoki:2016}.}.
More detailed studies including NN, $^3$He and $^4$He systems
are found in Ref.~\cite{Iritani:2016jie}.

\begin{table}
  \centering
  \begin{tabular}{cc|c|cc|c}
    \hline
    \hline 
    volume & $La$ & \# of conf. & \# of smeared sources  & $(A,B)$ & \# of wall sources \\
    \hline
    $32^3 \times 48$ & 2.9 fm & 402 & 384 & $(1.0, 0.18)$ & 48 \\
    $40^3 \times 48$ & 3.6 fm & 207 & 512 & $(0.8, 0.22)$ & 48 \\
    $48^3 \times 48$ & 4.3 fm & 200 & $4 \times 384$ & $(0.8, 0.23)$ & $4 \times 48$ \\
    $64^3 \times 64$ & 5.8 fm & 327 & $1 \times 256$ & $(0.8, 0.23)$ & $4 \times 64$ \\
    \hline
    \hline
  \end{tabular}
  \caption{Simulation parameters.
  The rotational symmetry for isotropic lattice is used to increase statistics.
  \label{tab:lat_param} }
\vspace {-2ex}
\end{table}

\vspace{-1.00ex}
\subsection{Quark source dependence of the HAL QCD potential}
\vspace{-1.00ex}

\begin{figure}[h]
  \centering
  \includegraphics[width=0.47\textwidth,clip]{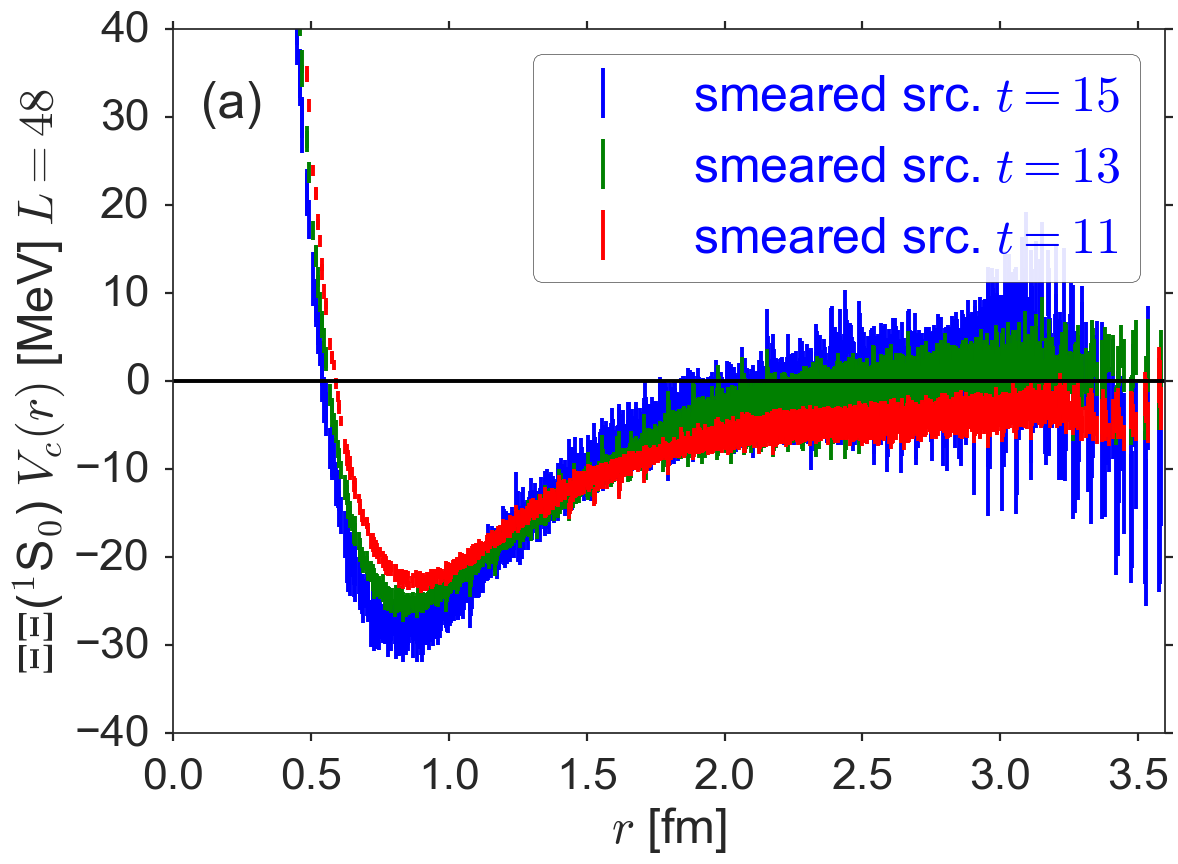}
  \includegraphics[width=0.47\textwidth,clip]{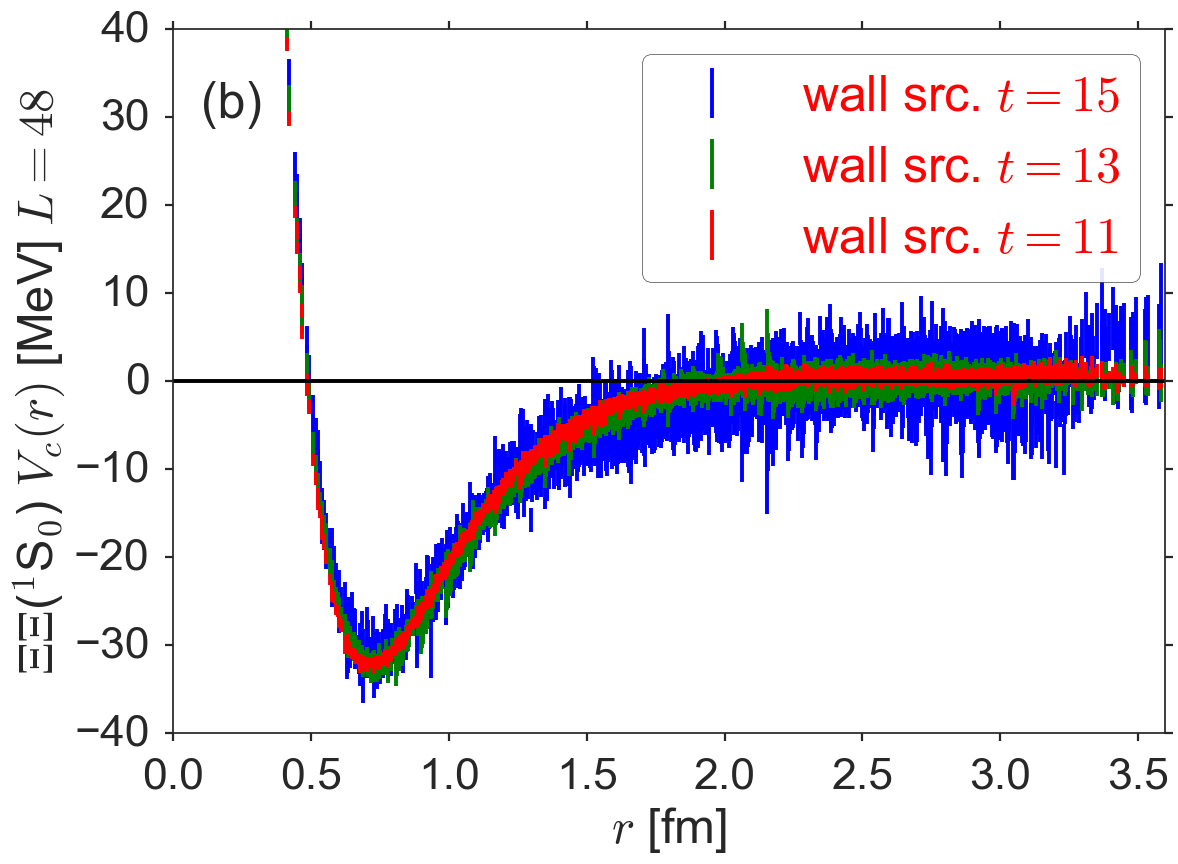}
  \includegraphics[width=0.47\textwidth,clip]{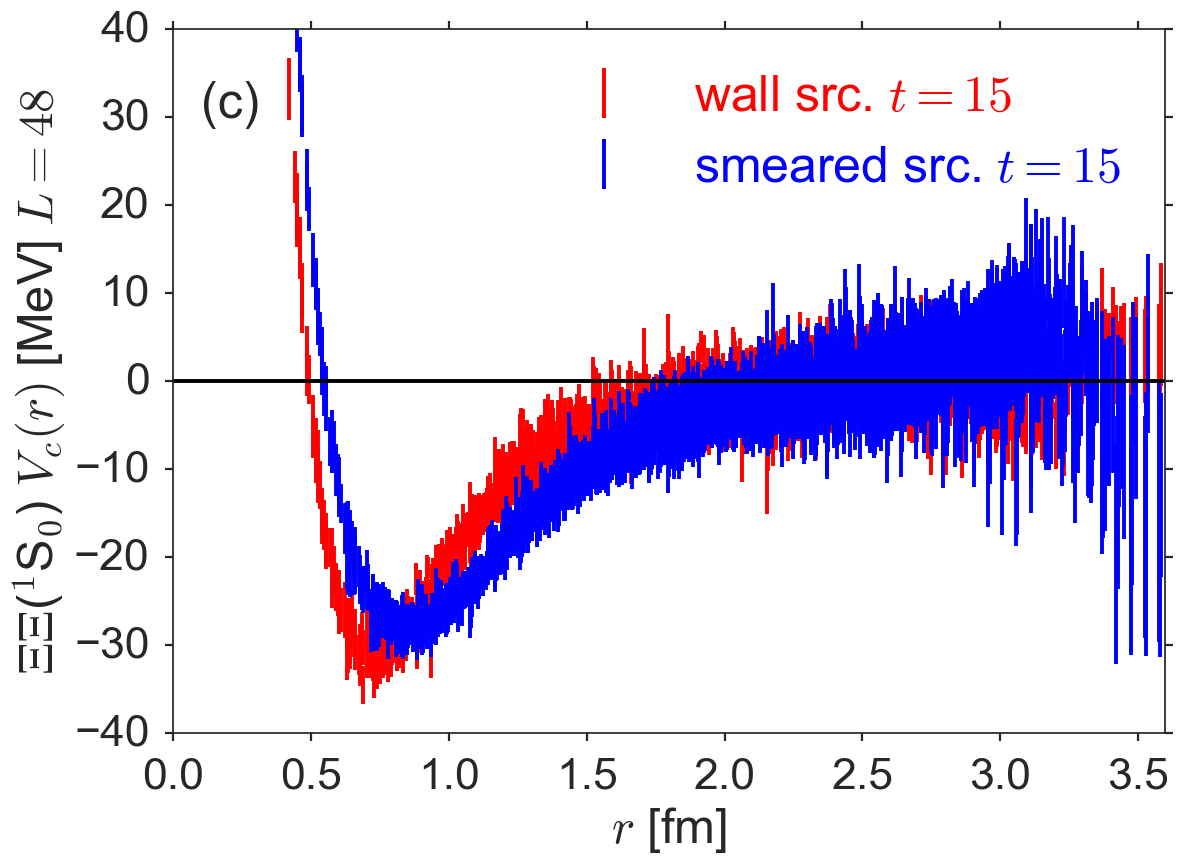}
  \includegraphics[width=0.47\textwidth,clip]{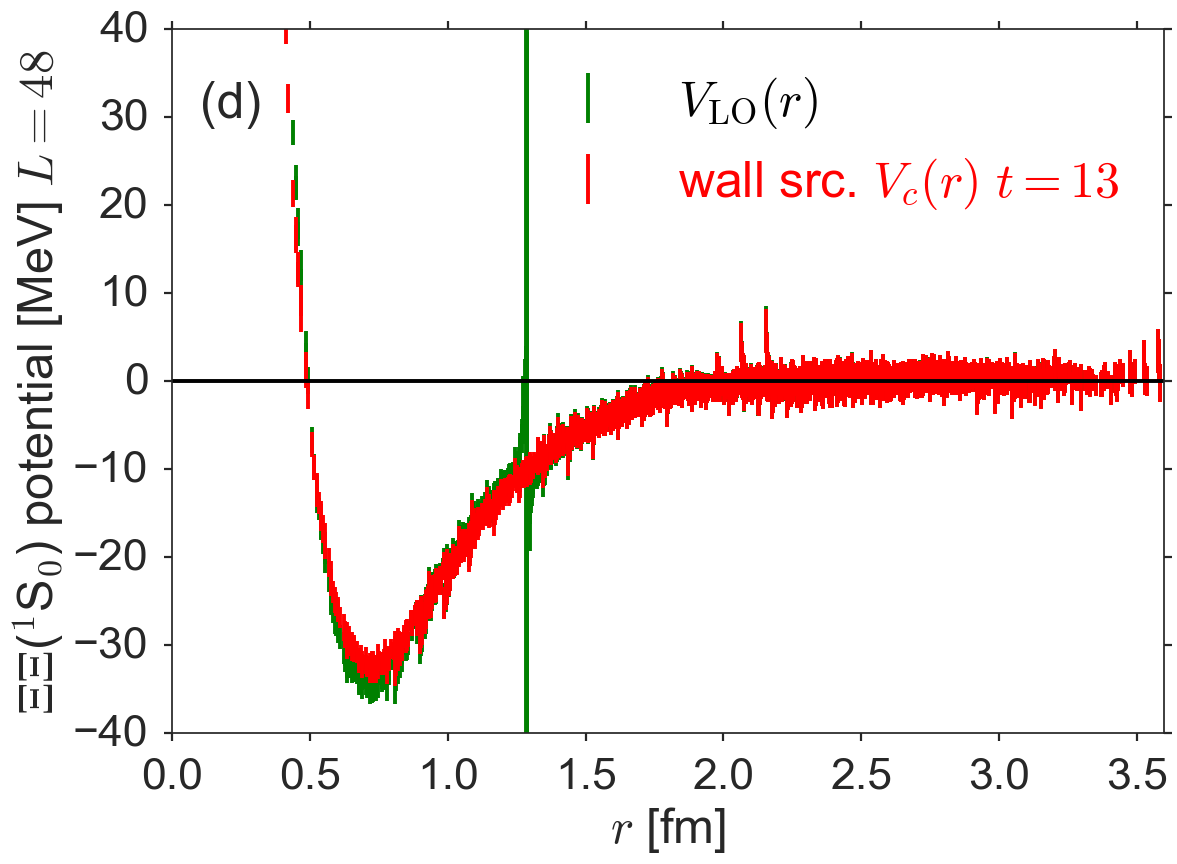}
  \caption{\label{fig:potentials} $V_C(r)$ of $\Xi\Xi$($^1$S$_0$) for $L^3 = 48^3$.
    (a) $V_C^\mathrm{smear}(r)$ at $t = 11,13$ and $15$ 
    (b) $V_C^\mathrm{wall}(r)$ at $t = 11$, $13$ and $15$.
    (c) a comparison between $V_C^\mathrm{wall}(r)$ and $V_C^\mathrm{smear}(r)$ at $t = 15$.
    (d) LO potential $V_\mathrm{LO}(r)$ and $V_{C}^\mathrm{wall}(r)$.}
\vspace{-2ex}
\end{figure}

We similarly consider the source dependence of the HAL QCD potential.
Fig.~\ref{fig:potentials}(a) and (b) show the central potential $V_C(r)$ of $\Xi\Xi$($^1$S$_0$) at $L^3 = 48^3$
from smeared and wall sources, respectively.
While $V_C^\mathrm{wall}(r)$ is stable against a variation of $t$ from $t = 11a$ to $15a$
within errors, 
$V_C^\mathrm{smear}(r)$ has a weak $t$ dependence and 
is slightly different from $V_C^\mathrm{wall}(r)$ as seen in Fig.~\ref{fig:potentials}(c) at $t = 15a$
though the difference decreases as $t$ increases.  

Contrary to the direct method, the source dependence in the HAL QCD method give an extra information, 
from which we can determine the next leading order of the derivative expansion as 
\begin{eqnarray}
  V^{X}_{C}(r)R^{X}(r,t) \equiv 
  \left[ \frac{1}{4m}\frac{\partial^2}{\partial t^2} - \frac{\partial}{\partial
  t} - H_0 \right]R^{X}(r,t)
  = V_\mathrm{LO}(r) R^{X}(r,t) + V_\mathrm{NLO}(r) \nabla^2 R^{X}(r,t) 
  \label{eq:NLOpotential}
\end{eqnarray}
with $X =$ wall, smeared.
As seen in Fig.~\ref{fig:potentials} (d),  $V_C^\mathrm{wall}(r)$ is a good approximation of 
$V_\mathrm{LO}(r)$, 
so that it gives reliable results at the low energy where $V_\mathrm{LO}(r)$ dominates.

\vspace{-1.00ex}
\subsection{Anatomy of fake plateaux by the potential } 
\vspace{-1.00ex}

\begin{figure}[h]
  \centering
  \includegraphics[width=0.47\textwidth,clip]{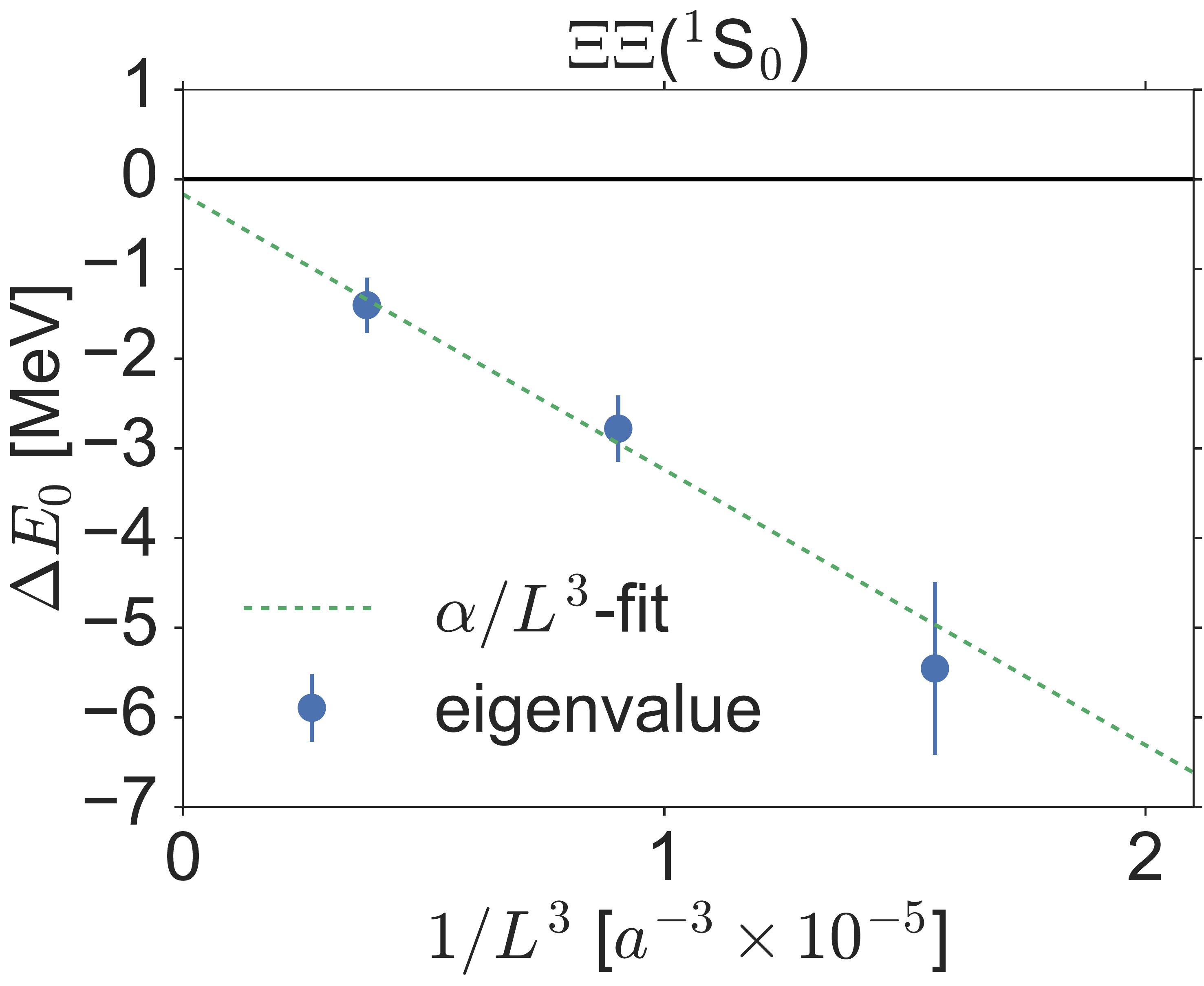}
  \includegraphics[width=0.47\textwidth,clip]{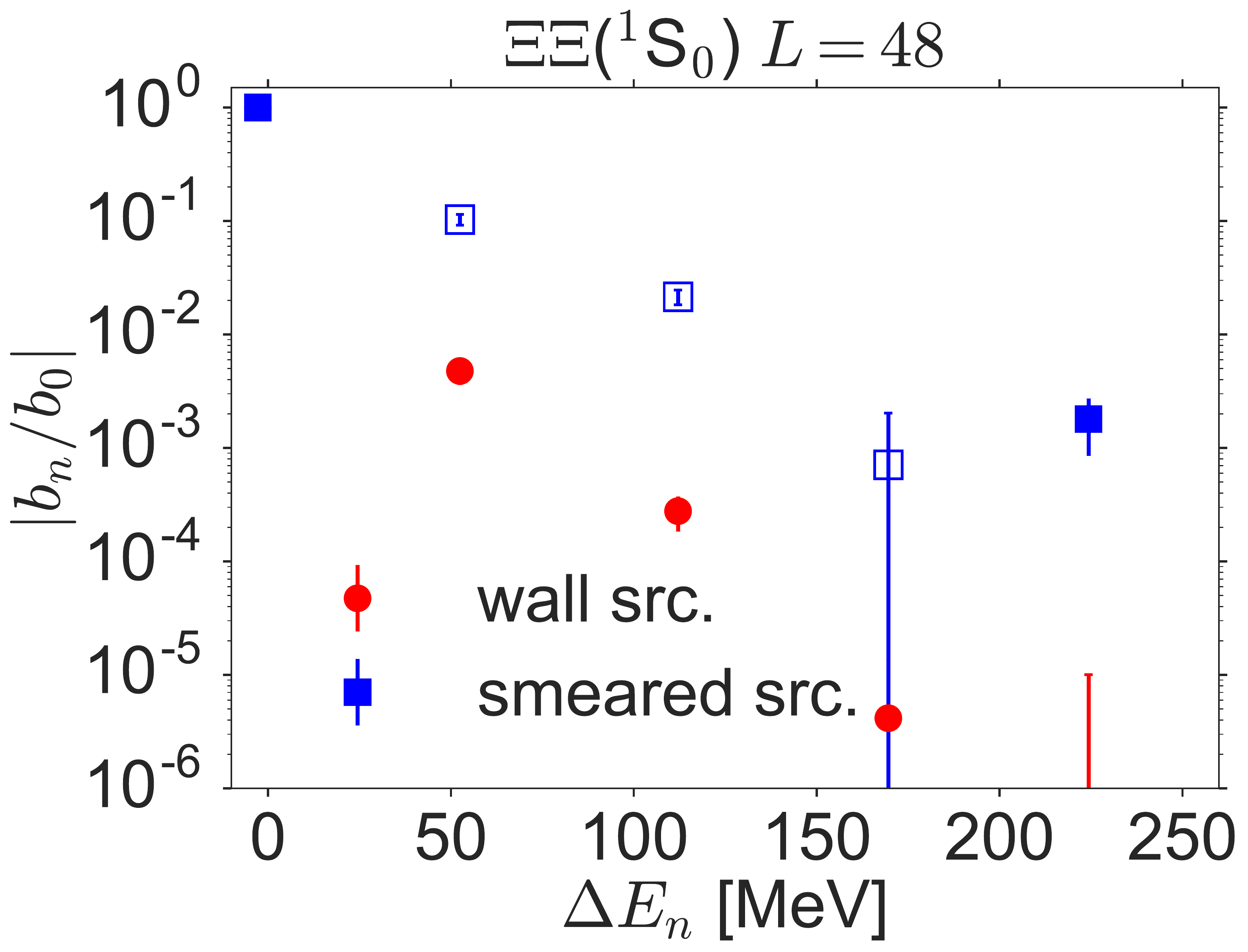}
  \caption{\label{fig:eigen} (a) The volume dependence of the ground state eigenvalue $\Delta E_0$.
    (b) The contamination of excited states $|b_n/b_0|$.  Solid (open) symbol denotes a positive(negative) value.}
\vspace{-2ex}
\end{figure}

While we have found no bound state in $\Xi\Xi(^1$S$_0$) channel 
from the Shr\"odinger equation with the HAL QCD potential in the infinite volume, 
eigenvalues of $H = H_0 + V$ on the finite volume $L$
gives the finite volume ground state energy \cite{Charron:2013paa,Iritani:2015dhu}. 
Fig.~\ref{fig:eigen}(a) shows the volume dependence of the lowest eigenvalue $\Delta E_0$
for $L^3 = 40^3, 48^3$ and $64^3$ from the wall source potential $V_C^\mathrm{wall}(r)$ at $t = 14a$
\footnote{The eigenvalues are consistent within errors from $t = 11a$ to $15a$.}, 
together with the linear extrapolation
 in $1/L^3$, which confirms the absence of the bound state in the $\Xi\Xi(^1$S$_0$) at $m_\pi = 0.51$ GeV.

Furthermore, using several low-lying eigenfunctions $\Psi_n(\vec{r})$ with eigenvalues $\Delta E_n$,
we can decompose $\Xi\Xi$ correlation functions as  
\begin{equation}
  \sum_{\vec r} R^\mathrm{wall/smear}(\vec{r},t) \simeq \sum_{\vec{r}} \sum_n a_n^\mathrm{wall/smear}
  \Psi_n(\vec{r}) \exp\left( -\Delta E_n t \right) =  \sum_n b_n^\mathrm{wall/smear} \exp\left( -\Delta E_n t \right),
  \label{eq:wave_decomposition}
\end{equation}
where $a_n^\mathrm{wall/smear}$ are determined from the orthogonality of $\Psi_n(\vec{r})$.
Fig.~\ref{fig:eigen}(b) shows the ratio $|b_n/b_0|$ 
as a function of the eigenvalue $\Delta E_n$, which shows
that the contamination of excited states. 
The contamination from the first excitation with about 50 MeV at $L^3 = 48^3$
is much smaller than 1\% for the wall source,
while it is about 10\% for the smeared source.

\begin{figure}[h]
  \vspace{-1.0ex}
  \centering
  \includegraphics[width=0.47\textwidth,clip]{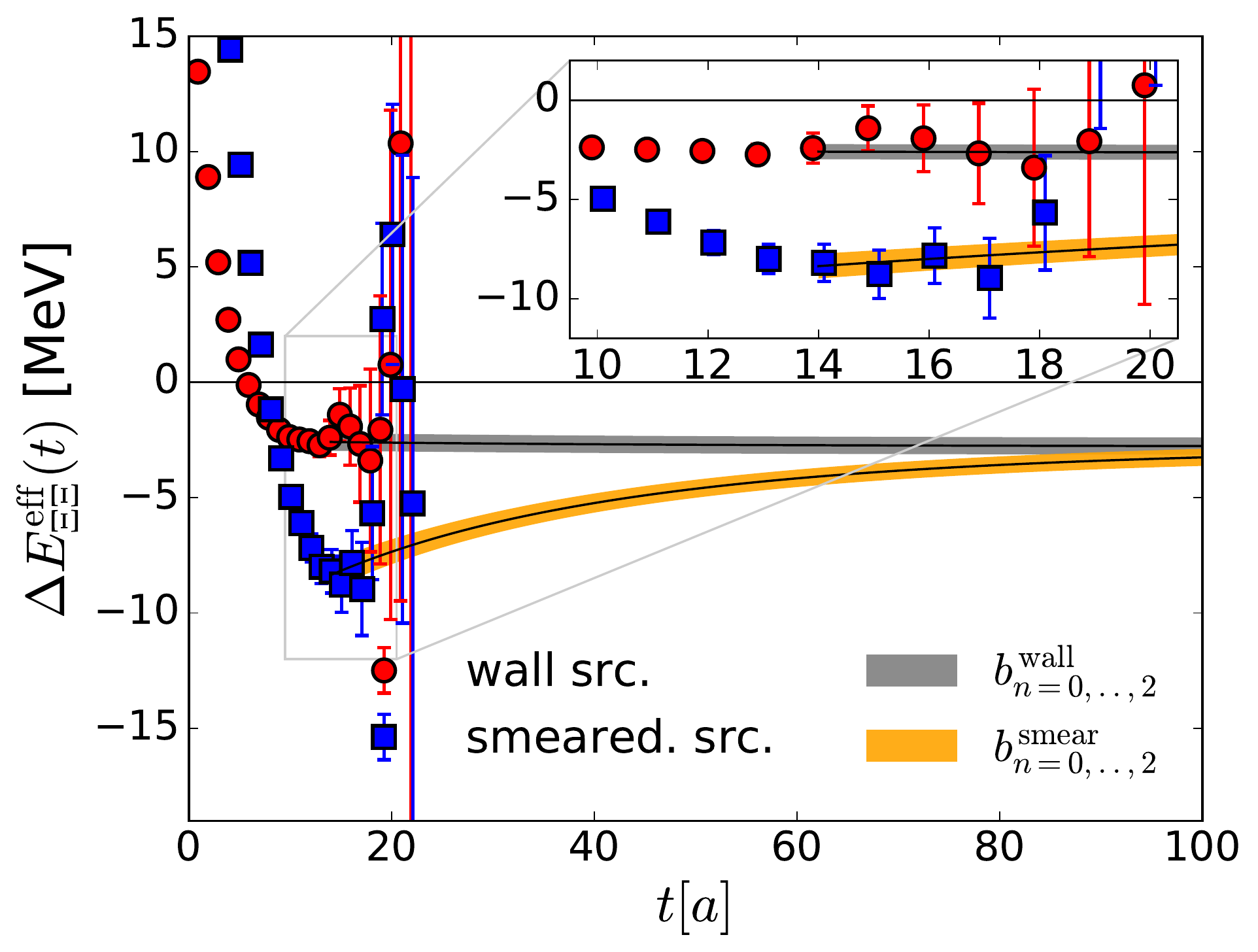}
 \includegraphics[width=0.46\textwidth,clip]{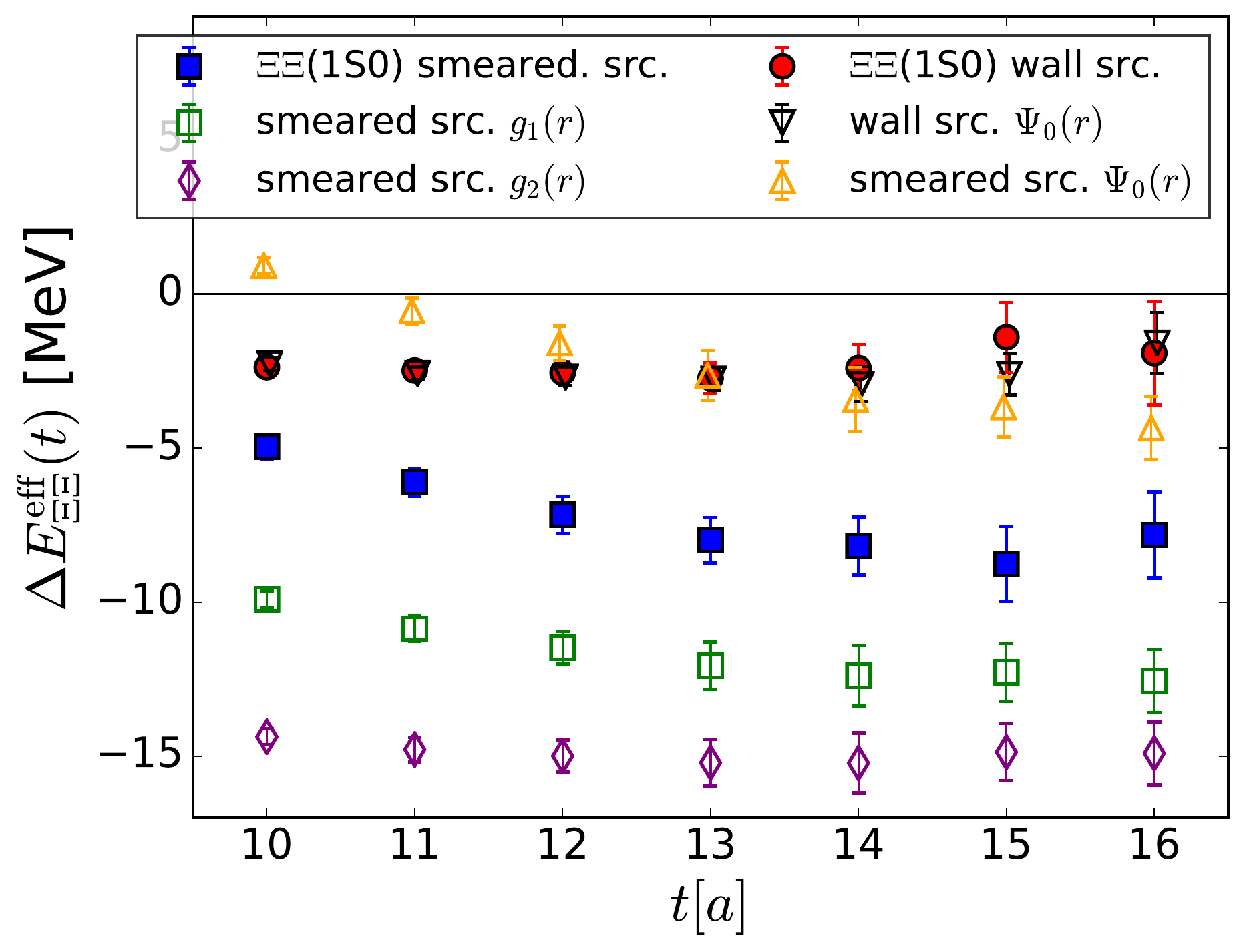}
  \vspace{-1.0ex}
  \caption{\label{fig:eff_reconst} (Left) Reconstructions of  $\Delta E_\mathrm{eff}(t)$ from  low-lying three eigenstates.
  (Right) Effective energy shift from sink projection by $g_1(r) =1-0.5 e^{-0.2 r}$, $g_2(r) = 1-0.9 e^{-0.22 r}$ and $\Psi_0$. }
  \vspace{-2.1ex}
\end{figure}

Using the decomposition Eq.~(\ref{eq:wave_decomposition}),
we can reconstruct the effective energy shift $\Delta E_\mathrm{eff}(t)$, as shown in 
Fig.~\ref{fig:eff_reconst} (Left), where the reconstructed result, denoted by  
the gray (orange) band for the wall (smeared) source is compared with the direct calculation.
The plateau-like structure for both sources is well explained by  
the reconstruction, while it is also shown that
the ground saturation for the smeared source requires $t \sim 100a \simeq 10$ fm \cite{Iritani:2016jie}.

The effective energy from
$\sum_{\vec r} g(r) R^{\rm smeared}(\vec r,t)$ is plotted in Fig.~\ref{fig:eff_reconst} (Right), 
which shows the strong sink operator dependence among $g(r)=1$ (solid square),
$g_1(r)$ (open square) and $g_2(r)$ (open diamond), while we confirm the
agreement among three for the wall source\cite{Iritani:2016jie}.  

Plateaux of the effective energy shift from $\sum_{\vec r} \Psi_0(\vec r)
R^{\rm wall/smeared}(\vec r,t)$, where $\Psi_0(\vec r)$ is the lowest
eigenstate at $t=14a$ on $L=48$, on the other hand, agree between the wall
(open down triangle) and the smeared (open up triangle) sources 
in  Fig.~\ref{fig:eff_reconst} (Right), 
where they also agree with that from the wall source without  $\Psi_0(r)$
(solid circle).
This analysis demonstrates that the lowest eigenstate from the potential is
indeed correct, and one can extract the correct lowest energy in the direct
method once we know the eigenstate. 
In the present case, the wall source happens to give the correct lowest energy
within errors in the direct method, though this is not always the case.

\vspace{2ex}

We have shown that the direct measurement for the energy shift has 
strong source and sink dependencies 
while the (time-dependent) HAL QCD method is free from these dependencies.
We also demonstrate that 
the origin of the fake plateau of the effective energy shift in the direct method can be clarified
by the lowest few eigenstates by using the potential on the finite volume. 

  \vspace{-2.00ex}
  We thank the authors of \cite{Yamazaki:2012hi} for providing
  the gauge configurations and
  the detailed account of the smeared source used in \cite{Yamazaki:2012hi}.
  The lattice QCD calculations have been performed on Blue Gene/Q at KEK 
  (Nos. 12/13-19, 13/14-22, 14/15-21, 15/16-12),
  HA-PACS at University of Tsukuba (Nos. 13a-23, 14a-20)
  and K computer at AICS (hp150085, hp160093).
  This research was supported by MEXT as ``Priority Issue on Post-K computer''
  (Elucidation of the Fundamental Laws and Evolution of the Universe)
  and JICFuS.

\end{document}